\documentclass[aps,pra,reprint,superscriptaddress]{revtex4-2}

\usepackage{amsmath,amssymb}
\usepackage{graphicx}
\usepackage{dcolumn}
\usepackage{bm}

\usepackage[colorlinks=true, allcolors=blue]{hyperref}
\usepackage{physics}
\usepackage{braket}
\usepackage[caption=false]{subfig}

\usepackage{xcolor}

\begin{document}

\title{Transition between one- and two-dimensional topology\\ in a Chern insulator of finite width}

\author{Frode Balling-Ansø}
\affiliation{Department of Physics and Astronomy, Aarhus University, DK-8000 Aarhus C, Denmark}

\author{Adipta Pal}
\affiliation{Max Planck Institute for Chemical Physics of Solids, N\"othnitzer Stra{\ss}e 40, 01187 Dresden, Germany}
\affiliation{Max Planck Institute for the Physics of Complex Systems,
N\"othnitzer Stra{\ss}e 38, 01187 Dresden, Germany}

\author{Ashley M. Cook}
\affiliation{Max Planck Institute for Chemical Physics of Solids, N\"othnitzer Stra{\ss}e 40, 01187 Dresden, Germany}
\affiliation{Max Planck Institute for the Physics of Complex Systems,
N\"othnitzer Stra{\ss}e 38, 01187 Dresden, Germany}

\author{Anne E. B. Nielsen}
\affiliation{Department of Physics and Astronomy, Aarhus University, DK-8000 Aarhus C, Denmark}

\begin{abstract}
Topology in quantum systems is typically considered in infinite crystals in one, two, or higher integer dimensions. Here, we show that one can continuously transform a system between a topological phase associated with one dimension and a topological phase associated with two dimensions without closing the energy gap. In this process, the dimension of the system itself changes. Concretely, we investigate a modified version of the Qi-Wu-Zhang model and develop a procedure to smoothly shrink the width of the system in one direction. By tracking gaps which remain open throughout the modulation, we establish a smooth transition from a two-dimensional to a one-dimensional topological insulator. In between the system exhibits both one- and two-dimensional topology, and the way the system accomplishes the transition is by making the one-dimensional topology more robust as the width decreases, while the two-dimensional topology becomes less robust. Finally, we show how the gaps arise from hybridization of edge states due to the finite width.
\end{abstract}

\maketitle

\section{Introduction}

The study of topological quantum matter provides an expansive framework for understanding phases of matter beyond its local microscopic properties. An important example of such is the study of topological insulators, where phases are described and distinguished using topological invariants for different symmetry classes and dimensions \cite{top1, Ryu_2010, top3, top2}. The classification of these phases is well-understood in the thermodynamic limit where the Brillouin zone turns continuous. This naturally begs the question of how the picture might change as we move away from the thermodynamic limit into systems with finite-size structures \cite{FiniteTop}. 

Moving to systems of spatially finite sizes allows for deviations from the usually robust properties of topological insulators. One of these properties is the cornerstone of the field, namely bulk-boundary correspondence \cite{BB1, BB2, BB3}. This states that a topologically non-trivial insulator subjected to periodic boundary conditions will exhibit conducting edge states as the boundaries are opened. However, finite system sizes allow for hybridization between the topological edge states thus turning the system insulating once again. This allows one to describe this new system as another topological insulator with another topological invariant. This has been observed in previous work \cite{FiniteTop} in which a modified version of the Qi-Wu-Zhang model \cite{QWZ} is shown to exhibit both one- and two dimensional topological signatures. This introduces phases of matter deviating from the usual description in the thermodynamic limit as they need to be described using two topological invariants. 

Later works report similar states in a wider variety of systems~\cite{calderon2023_fst, pacholski2023, pal2024_fstwsm}, in particular higher dimensional analogues of the earliest reported finite-size topological phases~\cite{FiniteTop}. These higher-dimensional analogues derive from intrinsically three dimensional topological states such as the strong topological insulator~\cite{PhysRevB.75.121306}, the Weyl semimetal~\cite{PhysRevLett.107.127205}, and the canonical crystalline topological insulator~\cite{PhysRevLett.106.106802} and occur in thin film geometries and therefore candidates for realisation in Van der Waals layered and heterostructure materials~\cite{vanwaals-hetero,Nature-hetero-antiferr,TI-hetero,robust-2dhetero, TMD-1,TMD-topo}. These analogues appear, at first glance, to exhibit bulk-boundary correspondence of two-dimensional topological states co-existing with response signatures of three-dimensional states, but these works instead revealed distinctions between the bulk-boundary correspondences of these states and those of known two-dimensional topological states, related to the dependence of this bulk-boundary correspondence on that of the underlying three-dimensional topological phase. These finite-size topological phases are now more deeply understood ---alongside two other sets of topological phases of matter~\cite{cook_multiplicative_2022, cook2023}---within the framework of the quantum skyrmion Hall effect~\cite{PhysRevB.109.155123, patil2024, patil2025microscopicfieldtheoriesquantum}. This framework reveals the significance of topological states in systems of spatially finite size as motivating generalisation of interpretation and treatment of internal state space associated with myriad (pseudo)spin degrees of freedom.  

We expand upon this investigation into lattice tight-binding Hamiltonians for Chern insulator states of spatially finite sizes by introducing a procedure to smoothly modulate the width of the system. This allows us to continuously move between the limiting cases of one- and two-dimensional system sizes. Throughout this modulation, we track different gaps and investigate how the topological signatures of different dimensions manifest and how robust these signatures are. Furthermore, we reproduce the observed gaps to high accuracy by computing the edge states in the thermodynamic limit and mapping them to finite-size systems. This further reinforces the fact that the gaps occur through hybridization under finite-size effects.

We find that the robustness of the topological signatures are inversely related. As we decrease the width of the system and move closer to a one-dimensional lattice, the one-dimensional signature becomes more robust. This is due to the gap sizes in the spectrum increasing, thus bracing the system against disorder. On the other hand, the two-dimensional signature becomes less robust as the topology is protected by real space distances between states in the system. By tracking gaps that remain open through the modulation, this establishes a smooth transition between one- and two-dimensional topological insulators. We find such gaps for cases of both trivial and non-trivial one-dimensional topology. In between these cases the system exhibits non-trivial topology with respect to both signatures at once. 


The article is structured as follows: In Sec.\ \ref{sec: Hamiltonian}, we describe the Hamiltonian as well as its properties. This entails describing its symmetries and deriving general properties of the spectrum. In Sec.\ \ref{sec:hybrid}, we calculate how the finite system size results in spectrum gaps through hybridization of edge states. In Sec.\ \ref{sec:modulation}, we then introduce a procedure to smoothly modulate the system width. In Sec.\ \ref{sec: robustness}, we define the one- and two-dimensional topological measures and apply them to determine how the topological properties of the system are affected by the smooth modulation of the system width. Section \ref{sec: Conclusion} concludes the paper. Appendix \ref{app_sec_A} provides further details of the computation of the hybridization due to finite witdh.

\section{Model}\label{sec: Hamiltonian}

In this section we summarize important properties of the modified Qi-Wu-Zhang model. This model, which was also considered in Ref.\ \cite{FiniteTop}, exhibits non-trivial two-dimensional topology, but as one of the boundaries is opened, the system remains insulating for finite system sizes, see Fig.\ \ref{fig:overlay_spectrum}. This allows the new system to be described using a one-dimensional topological invariant. As such, we get a single system exhibiting both one- and two-dimensional topological signatures. 

\subsection{The Hamiltonian}

The Hamiltonian we consider is a modified version of the Qi-Wu-Zhang model built on a two-dimensional lattice with $L_x$ and $L_y$ unit cells in each direction. Each unit cell has two internal degrees of freedom indexed by $\sigma \in \{1, -1\}$. For open boundary conditions in both spatial directions, the Hamiltonian is given by
\begin{align}
    H_{\textrm{oo}} = & \sum_{n=1}^{L_x}\sum_{m=1}^{L_y} \ket{n,m}\bra{n,m} \otimes (M\sigma_z) \\
    + & \sum_{n=1}^{L_x-1}\sum_{m=1}^{L_y} \ket{n,m}\bra{n+1,m} \otimes (-t\sigma_z-i\Delta \sigma_y)+h.c. \notag \\
    + & \sum_{n=1}^{L_x}\sum_{m=1}^{L_y-1} \ket{n,m}\bra{n,m+1} \otimes (-t\sigma_z-i\Delta \sigma_x) + h.c. \notag
\end{align}
where the index for $H_{\textrm{oo}}$ indicates open boundary conditions in both directions, and $\sigma_i$ denotes the usual Pauli matrices. $M$ denotes the on-site potential, and $t$ denotes the hopping amplitude between unit cells and the same internal sites, while $\Delta$ is the hopping between unit cells and different internal sites. Throughout the article, we pick $t=1$ and $\Delta = 0.22$ as well as setting $L_y = 500$ and $L_x$ to be even. 

We now impose periodic boundary conditions in the $y$-direction. Through translational invariance this block diagonalizes the Hamiltonian into momentum sectors of momentum $k_y = 2\pi q/L_y$ where $q \in \{0, 1, ... , L_y-1\}$. This yields the Hamiltonian
\begin{align}
    H_{\textrm{op}}(k_y) & =   \sum_{n=1}^{L_x} \ket{n}\bra{n} \otimes h(k_y)  \\
    & + \sum_{n = 1}^{L_x-1}\ket{n}\bra{n+1} \otimes (-t\sigma_z-i\Delta \sigma_y) + h.c. \notag
\end{align}
where
\begin{equation}
    h(k_y) = [M-2t\cos(k_y)]\sigma_z +2\Delta \sin(k_y) \sigma_x
\end{equation}
describes how the Hamiltonian acts within a single unit cell.

Finally, we can impose periodic boundary conditions in the $x$-direction as well which yields the Hamiltonian
\begin{multline}
    H_{\textrm{pp}}(k_x, k_y)  = 2\Delta \sin(k_y) \sigma_x +2\Delta \sin(k_x) \sigma_y \\ 
    + [M-2t\cos(k_x)-2t\cos(k_x)]\sigma_z.
\end{multline}

\subsection{Symmetries and Properties}\label{secsymprop}

\subsubsection{Inversion Symmetry}

We define an inversion symmetry operator $\mathcal{I}\sigma_z$ which acts on external degrees of freedom as a $\pi$ radians rotation around some central point and on the internal degrees of freedom by $\sigma_z$. We can thereby write its action on a basis element as
\begin{equation}
    \mathcal{I}\sigma_z \ket{n,m,\sigma} = \text{sgn}(\sigma)\ket{(L_x+1-n),(L_y+1-m),\sigma}.
\end{equation}
This is a symmetry for the total Hamiltonian for all three cases of boundary conditions. However, it maps momentum as $k_{x/y} \rightarrow -k_{x/y}$ which specifically leaves the $k_{x/y}=0, \pi$ spaces invariant. This enforces at least two-fold degeneracy for energies belonging to eigenvectors from non-invariant momentum spaces since if $\ket{\psi(k)}$ is an eigenvector with energy $E$ then $\mathcal{I}\sigma_z\ket{\psi(k)}$ is an eigenvector with the same energy belonging to a different momentum-space. On the invariant spaces translation and inversion commutes allowing simultaneous eigenstates. Since $(\mathcal{I}\sigma_z)^2 = \mathbb{I}$ it must have eigenvalues $+1$ ($-1$) which we refer to as positive (negative) parity. 

\subsubsection{Charge Conjugation}

We now discuss another symmetry of all models, namely charge conjugation $K\sigma_x$ where $K$ denotes complex conjugation. This operator anti-commutes with the Hamiltonian $H K\sigma_x = - K\sigma_x H$ which enforces symmetry in the spectrum around $E=0$ since if $\ket{\psi}$ is an eigenvector with energy $E$ then $ K\sigma_x \ket{\psi}$ is an eigenvector with energy $-E$. Furthermore, the two eigenvectors have opposite parity as inversion and charge conjugation anti-commute as well.

\subsubsection{General properties}

\begin{figure}
    \includegraphics[width=0.45\textwidth]{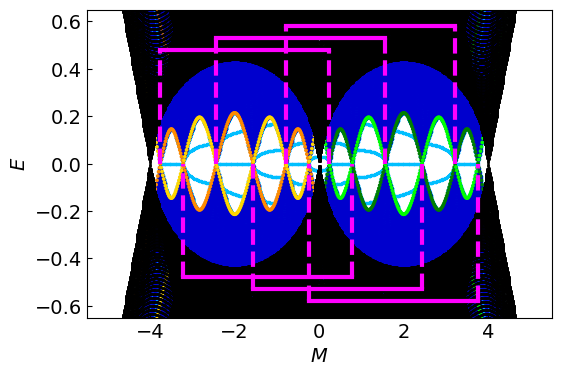}
    \captionsetup{justification=centerlast}
    \caption{The spectrum of $H_{\textrm{pp}}$ (black), $H_{\textrm{op}}$ (dark blue) and $H_{\textrm{oo}}$ (light blue)  for $L_x=6$. Six intervals of length $\Delta M = 4$ are shown in magenta with their edges coinciding with the gap closings. The spectrum of $H_{\textrm{op}}(k_y=0)$ is shown with positive (negative) parity states in light (dark) green. Likewise, the spectrum of $H_{\textrm{op}}(k_y=\pi)$ is shown with positive (negative) parity states in gold (orange). The spectra are symmetric around $E=0$ due to charge conjugation symmetry and around $M=0$ due to \eqref{uniequ}.}
    \label{fig:overlay_spectrum}
\end{figure}

Let us first consider the case of only periodic boundary conditions $H_{\textrm{pp}}$. The spectrum exhibits gap closings at $M=0, \pm 4$. Opening the boundary conditions in the $x$-direction yields the model $H_{\textrm{op}}$ which contains in-gap states in accordance with bulk-boundary correspondence. For large $L_x$, there seems to be no significant gaps in this spectrum. As one moves to shorter widths, however, gap openings become visible, and they become more pronounced the smaller $L_x$ is, see Fig.\ \ref{fig:overlay_spectrum}.

The spectrum of $H_{\textrm{op}}$ is symmetric around $M=0$ such that if there is an eigenstate at some $M$ with energy $E$ and momentum $k$ there must be an eigenstate at $-M$ with energy $E$ and momentum $k+\pi$. This is seen through unitary equivalence
\begin{equation}\label{uniequ}
    U \ H_{\textrm{op}}(M, k_y) \ U^{\dagger} = H_{\textrm{op}}(-M, k_y+\pi) 
\end{equation}
where $U$ is a unitary operator given by
\begin{equation}
    U = \sum_{n=1}^{L_x} (-1)^n\ket{L_x+1-n}\bra{n} \otimes \sigma_y.
\end{equation}

Any gap closing in the spectrum of $H_{\text{op}}(k_y)$ must occur at momentum $k_y = 0, \pi$. To see this, we may rewrite $H_{\text{op}}(k_y)$ as a sum of two Hermitian operators $H_1(k_y)=2\Delta \sin(k_y) \mathbb{I}_{L_x}\otimes \sigma_x$ containing only terms with $\sigma_x$ and $H_2(k_y)$ containing everything else. $H_{\text{op}}(k_y)$ has an eigenvalue of zero if and only if $H_{\text{op}}(k_y)^2$ has an eigenvalue of zero. Inspired by this, evaluating the square of the Hamiltonian yields
\begin{equation}
    H_{\text{op}}(k_y)^2 = (H_1+H_2)^2 = H_1^2+H_2^2+\{H_1, H_2\}.
\end{equation}
From the anti-commutation relations of Pauli operators it is evident that $\{H_1, H_2\}=0$ and we have $H_1^2 = 4 \Delta \sin^2(k_y) \mathbb{I}_{2L_x}$. As such, if $\ket{\psi}$ is an eigenvector of $H_2$ with eigenvalue $\lambda$ then it is also an eigenvector of $H_{\text{op}}(k_y)^2$ with eigenvalue $(4 \Delta \sin^2(k_y)+\lambda^2)$. This creates an absolute lower bound of eigenvalues $E(k_y)$ of $H_{\text{op}}(k_y)$ given by
\begin{equation}
    \abs{E(k_y)} \geq 2 \Delta \abs{\sin(k_y)},
\end{equation}
which immediately implies that gap closings only occur for $k_y = 0, \pi$.

As any gap closings occur for $k_y = 0, \pi$ we may investigate these momentum blocks to describe the nature of the gaps. We have the following property
\begin{equation}
    H_{\text{op}}(M,0) = H_{\text{op}}(M-4t, \pi).
\end{equation}
This means if $\ket{\psi}$ is a state that satisfies $H_{\text{op}}(0)\ket{\psi} = 0$ for some $M = M_0$ then it also satisfies $H_{\text{op}}(\pi)\ket{\psi} = 0$ for $M=M_0-4$ (as $t=1$). This means that all gap closings come in pairs with a mutual distance of $\Delta M = 4$. For a system width of $L_x$ the spectrum can be described as $L_x$ overlapping copies of these intervals. Then the edges of the intervals denote gap closings, see Fig.\ \ref{fig:overlay_spectrum}.

\section{Gap openings due to hybridization of Edge States}\label{sec:hybrid}

We now show explicitly that the gap openings arise from hybridization of the edge states. We provide the full calculation in the Appendix~\ref{app_sec_A} and only state the main results here. The Hamiltonian is chiral at the time-reversal-invariant momenta (TRIM) points $k_y=0,\pi$, and must therefore admit zero energy edge states for open boundaries in $x$. Here, we take the TRIM $k_y=0$ and linearize the Hamiltonian in the vicinity, which produces the left edge state in the semi-infinite limit, $x>0$. The position space wavefunction, $\psi_L(x)$, of the left edge state is localized near the boundary while decaying into the bulk ($x>0$) and must satisfy $\psi_L(x=0)=0$. Similarly, the right edge state can be derived in the semi-infinite limit, $x<L_x+1$. The position space wavefunction of the right edge state, $\psi_R(x)$, is localized at the boundary while decaying into the bulk ($x<L_x+1$) and must satisfy $\psi_R(L_x+1)=0$. 

If the system size in the $x$ direction is large, such that the left and right edge states do not overlap, the two semi-infinite limits can be treated independently and the effective surface Hamiltonian is given in the two-dimensional basis of the left and right edge states, denoted with the Pauli basis, $\nu$, as follows for $k_y$ small:
\begin{equation}
H_{0,\text{eff}}(k_y)=2\Delta k_y\nu_z.
\end{equation}

However, if the system size in the $x$ direction is smaller and comparable to the characteristic decay length scale of the edge states, such that there can be non-negligible wavefunction overlap, then the above surface Hamiltonian is not diagonal and the off-diagonal elements are derived from overlap terms. In that case, the effective surface Hamiltonian, still expressed in the semi-infinite limit edge state basis, is modified as follows:
\begin{equation}
\begin{split}
&H_{\text{eff}}(k_y)=2\Delta k_y\nu_z+\delta\nu_x,\\ 
&\delta = (t+\Delta)\psi_R^*(L_x)\psi_L(L_x+1).
\end{split}
\end{equation}
Non-negligible hybridization between the edge states, represented by $\delta$, yields a finite gap in the spectrum at $k_y=0$. The expression of $\delta$ shows that it can be zero if the left edge state satisfies a further boundary condition, $\psi_L(x=L_x+1)=0$, yielding a gapless spectrum in the $k_y=0$ sector. This is possible when $t>\Delta$, so that each of the left and right edge states transition from overdamped to damped-oscillatory position space wavefunctions. In the small system size regime, the left edge state satisfies the second boundary condition when its oscillatory component is a standing wave in the particle-in-a-box problem of size $L_x+1$. This explanation similarly carries for the right edge state with a second boundary condition, $\psi_R(x=0)=0$.

The wavelength of the oscillatory part of each edge state is parametrized by $M$, $t$ and $\Delta$ so that it can be shown that the gapless behavior at $k_y=0$ is regained when the following relation is satisfied,
\begin{equation}
M-2t=2\sqrt{t^2-\Delta^2}\cos\bigg{(}\frac{n\pi}{L_x+1}\bigg{)},\quad (n=1,...,L_x).
\end{equation}
Here $\frac{n\pi}{L_x+1}$ accounts for the wavevector of the standing wave. We show the detailed calculation, including expressions for the left and right edge wavefunctions, as well as the energy eigenvalues of $H_{\text{eff}}(k_y)$ vs.\ $M$ compared with numerics in Appendix~\ref{app_sec_A}.

\section{Smooth Modulation of System Width}\label{sec:modulation}

Our main area of interest is how the topological signatures depend on the width $L_x$ of the system. As such, we implement a way to smoothly modulate the width by introducing an orbital dependent potential along the edges. For the Hamiltonian $H_{\textrm{op}}(k_y)$ we thereby add the extra term
\begin{equation}
    H_{\text{edge}}(u)=\left(\ket{1}\bra{1}+\ket{L_x}\bra{L_x}\right) \otimes\tan\left(\frac{\pi u}{2}\right)\sigma_z
\end{equation}
where $u\in [0,1)$ is the parameter used to modulate the strength of the potential. At $u = 0$ the potential disappears and when $u \rightarrow 1$ it diverges. The $\sigma_z$ ensures half-filling on the edges as $u \rightarrow 1$. Note also that all the symmetries discussed in section \ref{secsymprop} are still present when adding $H_{\text{edge}}(u)$ to $H_{\textrm{op}}$, except the one in Eq.\ (\ref{uniequ}).

When the edges are frozen out due to the high potential, the remaining part of the system is effectively $H_{\textrm{op}}(k_y)$ but for a width of $L_x-2$ which has been verified numerically. We can then repeat this process by cutting away the frozen edges and implementing $H_{\text{edge}}(u)$ on the new edges to further decrease the system width. This results in a procedure that allows for a smooth change in $L_x$.

\begin{figure}
    \includegraphics[width=0.45\textwidth]{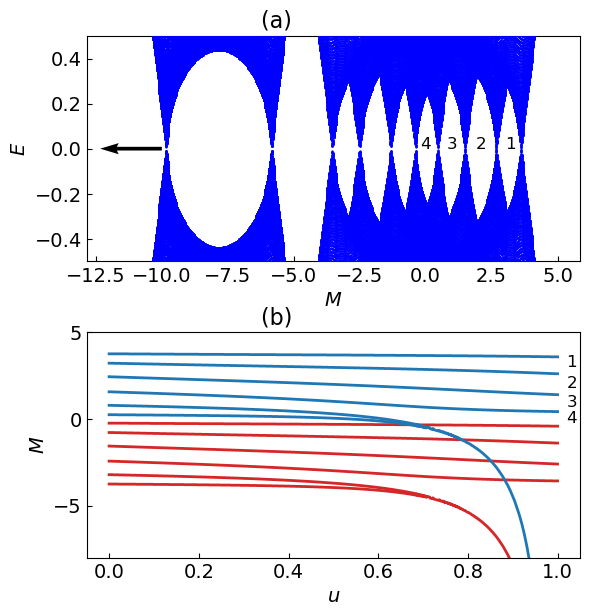}
    \captionsetup{justification=centerlast}
    \caption{A visualization of the smooth width modulation of the spectrum for $L_x=6$. (a) A snapshot of the spectrum for added edge potential $H_{\text{edge}}(u)$ with $u=0.918$. Two of the $\Delta M = 4$ intervals merge while they are shifted to the left away from the rest of the spectrum. The remaining part forms the $L_x=4$ spectrum. We label the four rightmost gaps from one to four for later reference as they never close. (b) The gap closings of the spectrum throughout the modulation. Gaps closings for $k_y=0$ ($k_y=\pi$) are highlighted in blue (red). Here gabs one through four can be seen to remain open throughout the whole process.}
    \label{fig:edge}
\end{figure}

As we increase $u$ and thereby the edge potential, two of the intervals should disappear to match the final total spectrum. Numerically, we observe that the two leftmost intervals start to perfectly overlap and are shifted to the left with increasing magnitude, see Fig.\ \ref{fig:edge}. Once the frozen edges are cut, they disappear completely. This is in accordance with expectation, as $H_{\text{edge}}(u)$ enters two of the single site Hamiltonians $h(k_y)$ as a shift in the negative $M$ direction. As such, the  $L_x-2$ rightmost gaps never close as the edge potential is increased.

\section{Analysis of Topological signatures}\label{sec: robustness}

To study the effect of system width on topological signatures, we smoothly shrink the system from $L_x=20$ to $L_x=6$. As seen in Fig.\ \ref{fig:edge}, the rightmost gaps remain open throughout the modulation. As such, we track the center of the four rightmost gaps and number them from the highest value of $M$ to the lowest (right to left). For every step in this process, we apply topological measures to investigate how they might be affected.

\subsection{One-dimensional Topology: The Wilson Loop}

\subsubsection{Definition}

We now consider a one-dimensional topological measure, which is based on the Wilson loop. The Wilson loop stems from parallel transporting the occupied states at a given momentum around in a loop in the Brillouin zone \cite{BerryPhase,Altland_Simons_2023}. This yields a gauge transformation whose matrix elements are given by
\begin{equation}
    W_{ij} = \bra{\psi^0_i} P_{\text{occ}}^{L_y-1} P_{\text{occ}}^{L_y-2} ... P_{occ}^{2} P_{\text{occ}}^{1}\ket{\psi^0_j}
\end{equation}
where $P_{\text{occ}}^{q}$ denotes the projector onto the subspace of occupied states with momentum $k_y = 2\pi q/L_y$, and $\{ \ket{\psi^0_j} \}$ denotes the basis of occupied states belonging to the subspace of momentum $k_y = 0$. 

If the system is in a trivial phase, the Wilson loop can be adiabatically deformed to the identity. Thus, topological non-triviality occurs if the matrix has topologically protected eigenvalues different from one. This is presented in Ref.\ \cite{Wilson} which derives a formula for the number of topologically protected eigenvalues of $-1$ denoted $N_{(-1)}$ for systems subject to inversion symmetry. $N_{(-1)}$ is derived from the number of negative parity states in the occupied inversion-invariant momentum subspaces
\begin{equation}
    N_{(-1)} = \abs{n_{-}(0)-n_{-}(\pi)}
\end{equation}
where $n_{-}(k_y)$ denotes the number of occupied negative parity states with momentum $k_y$. For at full derivation of this formula see Ref.\ \cite{Wilson}.

For computational ease we may write $N_{(-1)}$ in terms of basis invariant quantities. Let $\mathcal{I}_{k_y}$ denote the inversion operator $\mathcal{I}\sigma_z$ restricted to each of the occupied momentum subspaces. We may then write
\begin{equation}
    n_-(0) = \frac{1}{2}(L_x-\Tr(\mathcal{I}_{0}))
\end{equation}
Which yields the expression
\begin{equation}
    N_{(-1)} = \frac{1}{2} \abs{\Tr(\mathcal{I}_{0})-\Tr(\mathcal{I}_{\pi})}
\end{equation}
%

\subsubsection{Signature}

\begin{figure}
    \includegraphics[width=0.45\textwidth]{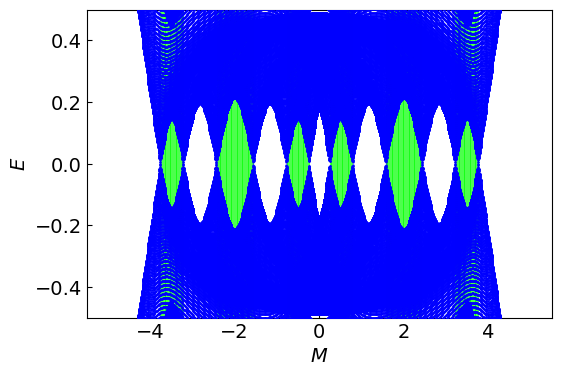}
    \captionsetup{justification=centerlast}
    \caption{The spectrum of $H_{\textrm{op}}$ for $L_x=6$. The gaps are colored according to the Wilson topological signature $N_{(-)}$ with trivial $N_{(-)}=0$ gaps depicted in white and non-trivial $N_{(-)}=1$ gaps depicted in green.}
    \label{fig:wilson}
\end{figure}

We start by applying the Wilson loop signature to the system with both open and periodic boundary conditions $H_{\textrm{op}}$. This shows that consecutive gaps alternate between having non-trivial and trivial topology, see Fig.\ \ref{fig:wilson}. We may understand this using the previously mentioned notion of intervals of $\Delta M = 4$. For each of these intervals we assign the value $1 \in \mathbb{Z}_2$. Picking a gap in the total spectrum, we can count how many intervals it lies within and sum up the corresponding numbers in $\mathbb{Z}_2$. If this number is $0 \in \mathbb{Z}_2$ the topology is trivial and if it is $1 \in \mathbb{Z}_2$, then the topology is non-trivial. 

\subsubsection{Robustness}

\begin{figure}
    \includegraphics[width=0.45\textwidth]{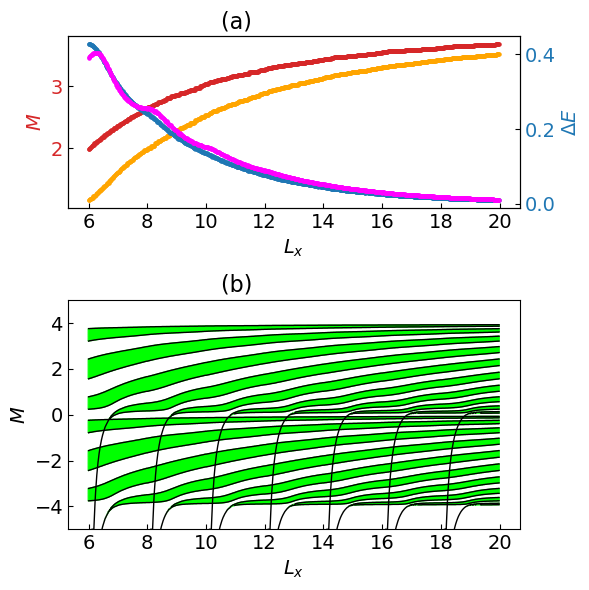}
    \captionsetup{justification=centerlast}
    \caption{Visualization of gaps in the spectrum throughout the smooth modulation of the width. (a) The size of a topological (non-topological) energy gap shown in blue (purple) at the $M$ value shown in red (orange) as it is tracked throughout the smooth width modulation. This shows that for smaller values of $L_x$ gaps become more prominent. (b) Gaps are shown throughout the modulation with topological (non-topological) gaps highlighted in green (white). The black lines show the gap closings. It is seen that the top six gaps remain open throughout the whole modulation.}
    \label{fig:gap_size}
\end{figure}

For the case of the Wilson loop signature, we know that any topological gaps remain topological throughout the whole process of shrinking the system width. The robustness of this signature increases for lower values of $L_x$ as the energy separation between the highest occupied state and the lowest unoccupied state increases. This is illustrated in Fig.\ \ref{fig:gap_size} which shows both topological and non-topological gaps. This implies that the one dimensional topological signature becomes more robust as we approach a spatially one-dimensional system.

\subsection{Two-dimensional Topology: The Adiabatic Charge Pump}

The two-dimensional topological measure is based upon the Chern number which is usually calculated as an integral of the Berry curvature over the two-dimensional Brillouin zone \cite{Altland_Simons_2023, intro_top}. This is, however, only possible for the case of periodic boundary conditions in both spatial directions. As such we need another way to compute it for other boundary conditions, and to this end we apply the adiabatic charge pump \cite{Tong, pump}. 

\subsubsection{Definition}

\begin{figure}
    \includegraphics[width=0.45\textwidth]{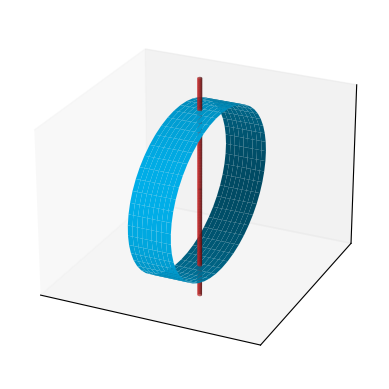}
    \captionsetup{justification=centerlast}
    \caption{Visual representation of how a flux-tube (red) is threaded though the periodic crystal (blue) in a way that preserves inversion symmetry.}
    \label{fig:flux_tube}
\end{figure}

\begin{figure*}[ht]
  \centering
  \captionsetup[subfloat]{position=top}

  \subfloat[]{  
    \includegraphics[width=0.4\textwidth]{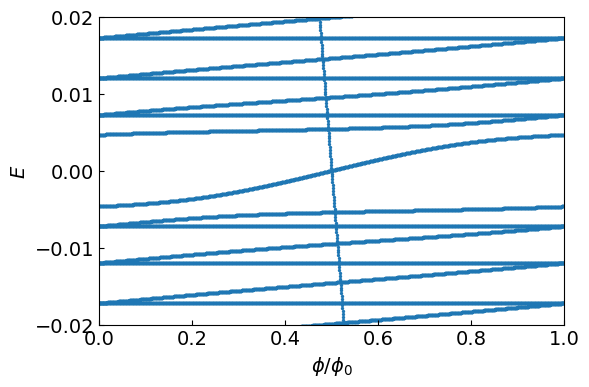}%
    \label{fig:op_charge}%
    }%
  \subfloat[]{%
    \includegraphics[width=0.4\textwidth]{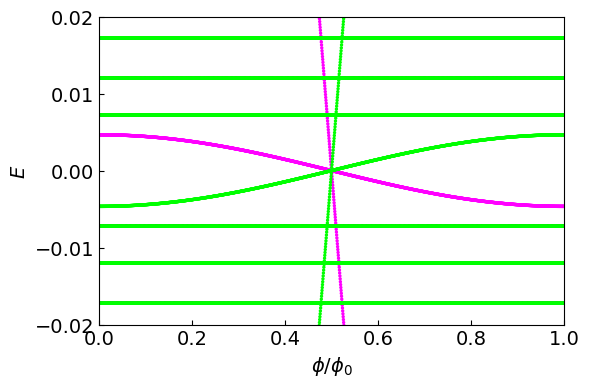}%
    \label{fig:op_charge_inv}%
  }
  \vfill
    \subfloat[]{%
    \includegraphics[width=0.4\textwidth]{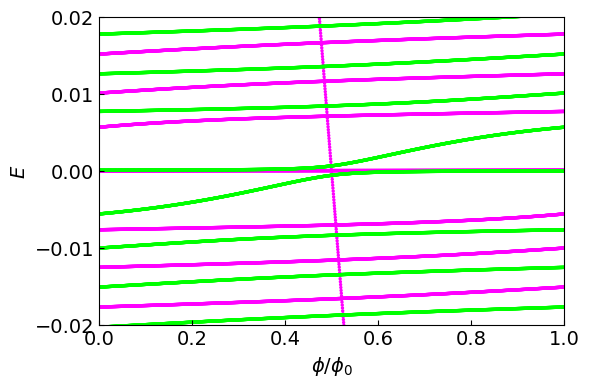}%
    \label{fig:oo_charge}%
  }
  \subfloat[]{  
    \includegraphics[width=0.4\textwidth]{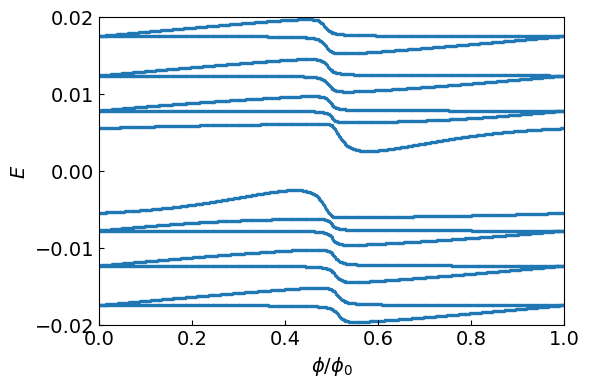}%
    \label{fig:op_charge_nontop}%
    }%

\caption{Visualization of the adiabatic charge pump with $L_x=20$. For cases preserving inversion symmetry positive (negative) parity states are depicted in green (magenta). Panels (a), (b) and (c) are computed for gap number three ($M=3.68$), while (d) is for number four ($M=3.52$). The graphs are computed from (a) $H_{\textrm{op}}$ with the flux-tube penetrating a single plaquette. This configuation breaks inversion symmetry but implies a Chern number of $C=1$. (b) $H_{\textrm{op}}$ with the flux-tube penetrating opposite ends of the periodic band. This shows transport as in (a) but now with two opposite contributions present. (c) $H_{\textrm{oo}}$ showing the same kind of transport as in (a), however, now there are zero energy states, which hybridise with the transported state. (d) $H_{\textrm{op}}$ through a single plaquette for a non-topological gap. Although avoided crossings are present throughout the spectrum, there is no transport past the Fermi energy.}
\label{fig:charge_pump}
\end{figure*}

\begin{figure*}[ht]
\captionsetup[subfloat]{position=top}
\centering
  \subfloat[]{  
    \includegraphics[width=0.35\textwidth]{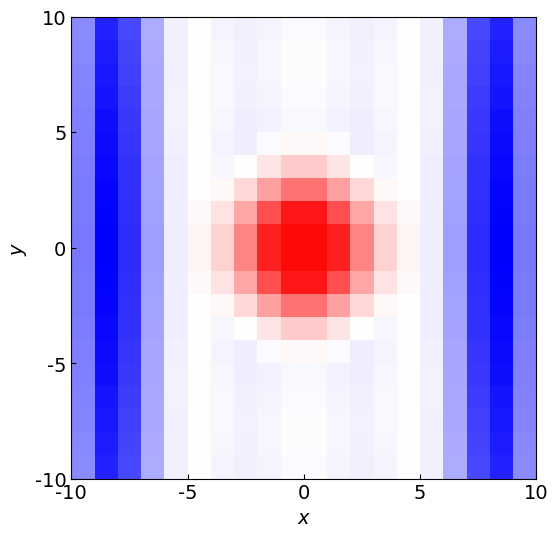}%
    \label{fig:density1}%
    }%
  \subfloat[]{%
    \includegraphics[width=0.35\textwidth]{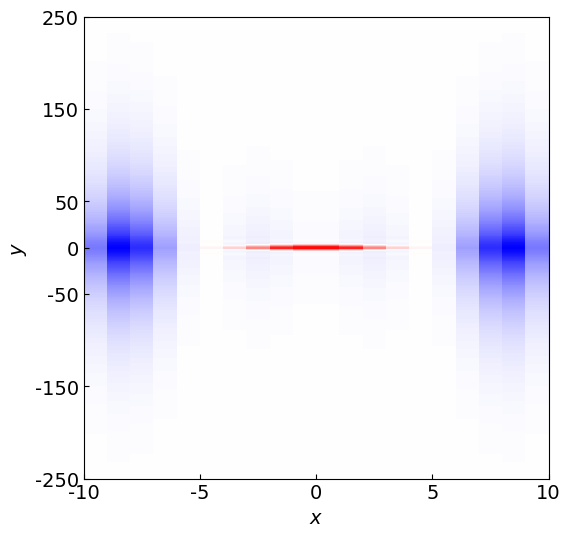}%
    \label{fig:density2}%
  }
\caption{Density plot showing the localization of the two states by the crossing ($\phi/\phi_0 = 0.498$ and $M = 3.68$) for $H_{\textrm{op}}$ and $L_x=20$. The graphs are computed for the inversion symmetric adiabatic charge pump from Fig.\ \ref{fig:op_charge_inv}. The bulk state is highlighted in red while the edge state is blue. (a) A zoomed in view around the location of the flux-tube with equal length scales on both axes. (b) A graph of the states over the whole crystal.}
\label{fig:density}
\end{figure*}

We first consider the model with open boundary conditions in both directions $H_{\textrm{oo}}$. We can think of this model as being embedded in a two-dimensional plane within three-dimensional space, say, the $xy$-plane. We now introduce a point-localized flux-tube through one of the plaquettes with flux $\phi(t) = \phi_0 \tau/T$ where $\phi_0 = 2 \pi$ is the flux quantum and $\tau \in [0, T]$ is a parameter used to modulate the flux, while $T$ is a scale chosen such that the modulation is adiabatic at all times. The flux-tube is placed in the center plaquette to preserve inversion symmetry of the model. This induces a magnetic potential which in cylindrical coordinates with respect to the flux-tube is given by $\textbf{A} = \frac{\tau}{rT} \hat{\phi}$. This vector potential manifests within our model as a Peierls phase
\begin{equation}
    \ket{\textbf{r}_j}\bra{\textbf{r}_i} \rightarrow \ket{\textbf{r}_j}\bra{\textbf{r}_i} e^{i\theta_{ij}} 
\end{equation}
where $\textbf{r}_j$ denotes the coordinate of a given site, and the Peierls phase $\theta_{ij}$ is given by
\begin{equation}
    \theta_{ij} = \int_{\textbf{r}_i}^{\textbf{r}_j} \textbf{A}(\textbf{r}) \cdot d\textbf{r}.
\end{equation}
To calculate the phase one can use independence of integration path following from Stoke's theorem. Introducing the complex coordinates $z_j = x_j+iy_j$ to denote the position of the vector $\textbf{r}_j$ relative to the flux-tube we can write the Peierls phase as
\begin{equation}
    \theta_{ij} = \frac{\tau}{T} \arg \left( \frac{z_j}{z_i} \right)
\end{equation}
where the $\arg$-function maps into the interval $[-\pi, \pi)$. We can simplify this within our model using the gauge transformation $\ket{\textbf{r}_j} \rightarrow \ket{\textbf{r}_j} \exp(-i\frac{\tau}{T}\arg(z_j))$. This cancels the Peierls phase in most places except for hoppings across a line drawn from the flux-tube outward in the negative $x$-direction. The final phase for these hoppings is given by $\exp \left(\pm 2 \pi i \tau/T\right)$ with the sign depending on the direction in which the line is crossed. From this, one sees that the spectrum is the same at the beginning and end of the flux-modulation. During the process, however, states may be pumped across the gap between occupied and unoccupied states. The Chern number is then given by the number of edge states pumped across the gap and the sign is given by the direction in which these states are pumped. 

We now consider the model with open and periodic boundary conditions $H_{\textrm{op}}$. For this model we can no longer embed the crystal in a two-dimensional plane in a way that preserves inversion symmetry. As such one should geometrically think of it as a band in three-dimensional space. If the flux-tube penetrates one plaquette, it must therefore also penetrate the one on the opposite side, see Fig.\ \ref{fig:flux_tube}. This creates two different transport effects in opposite directions. Nevertheless, one may still determine the Chern number through counting the number of edge states pumped across the gap in a given direction.

If we just thread the flux-tube through a single plaquette as in the case of $H_{\textrm{oo}}$ then inversion symmetry is broken. However, close to the flux tube the difference is numerically insignificant and inversion symmetry is approximately preserved. As we only interest ourselves in the area close to the flux-tube we may also apply this method without significant error. 

\subsubsection{Signature}

Applying the adiabatic charge pump shows that gaps identified by the Wilson loop as being topological also shows non-trivial charge transport, see Fig.\ \ref{fig:charge_pump}. For positive $M$ these gaps have a Chern number of $C=+1$ while for negative $M$ we have $C=-1$. Gaps identified by the Wilson loop as trivial shows no transport (Fig.\ \ref{fig:op_charge_nontop}) such that $C=0$.

Computing the adiabatic charge pump for completely open boundary conditions $H_{\textrm{oo}}$ yields the same results, but with zero energy edge states present in the spectrum. These states experience no transport during the flux modulation, which verifies that they correspond to a different topological signature, namely, the Wilson loop.

\subsubsection{Robustness}

\begin{figure*}[ht]
  \centering
  \captionsetup[subfloat]{position=top}

  \subfloat[]{  
    \includegraphics[width=0.4\textwidth]{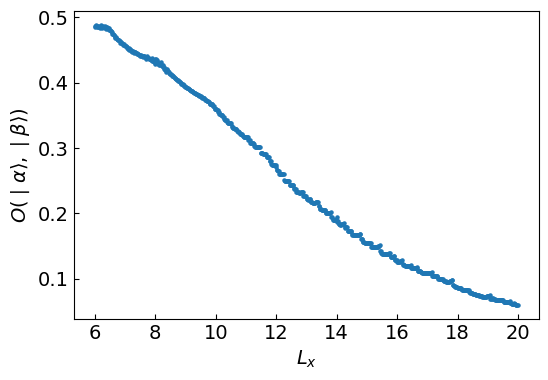}%
    \label{fig:op_overlap}%
    }%
  \subfloat[]{%
    \includegraphics[width=0.4\textwidth]{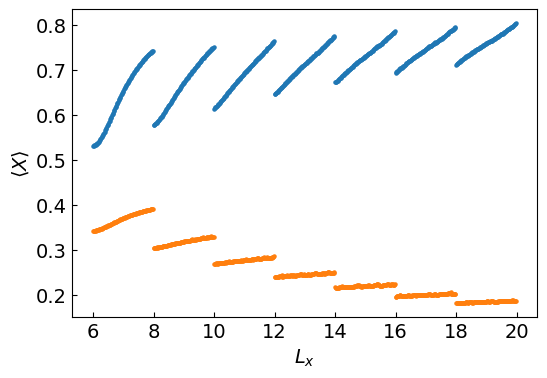}%
    \label{fig:op_x}%
  }
  \vfill
  \subfloat[]{  
    \includegraphics[width=0.4\textwidth]{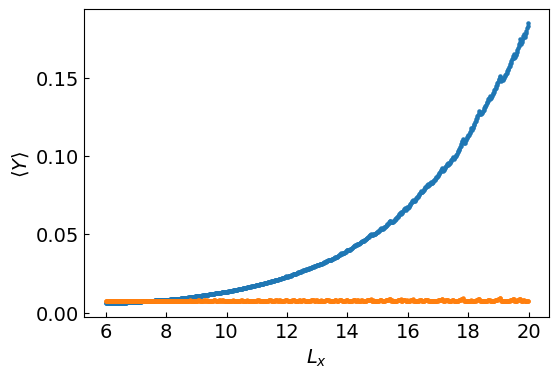}%
    \label{fig:op_y}%
    }%
  \subfloat[]{%
    \includegraphics[width=0.4\textwidth]{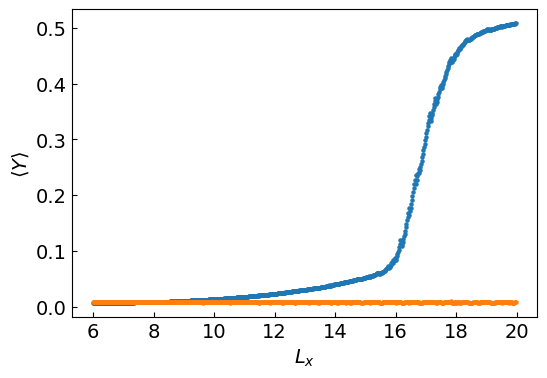}%
    \label{fig:oo_y}%
  }
\caption{Graphs of the spatial measures computed over a modulation of the system width. All the figures are computed for gap number three starting at $M=3.68$ for $L_x=20$. (a) The density overlap measure $O(\ket{\alpha}, \ket{\beta})$ from Eq.\ (\ref{eq:overlap_measure}) calculated for the two states by the crossing in Fig.\ \ref{fig:op_charge_inv} for the system $H_{\textrm{op}}$. This shows lower overlap for larger widths which serves to protects robustness of the crossing. (b) Calculation of $\langle X \rangle$ for the two states from (a). This shows one state moving closer into the bulk, while the other moves towards the edge. Discontinuities are due to the change in normalization for $X$. (c) Calculation of $\langle Y \rangle$ for the two states from (a). One moves exponentially away from the flux-tube, while the other stay by it. (d) Calculation of $\langle Y \rangle$ as in (c) but for completely open boundary conditions $H_{\textrm{oo}}$. This shows a sharp increase in $\langle Y \rangle$ due to hybridization with the zero energy edge states.}
\label{fig:real_space}
\end{figure*}

For the case of adiabatic charge transport, we need to check how robust the transport across $E=0$ is. To do this, we investigate the energy crossing in the gap in terms of how easily the two corresponding states may hybridize and create an avoided crossing. The crossing is at $\phi = \phi_0/2$ but to keep the states non-degenerate, we analyze them at $\phi = 0.498 \phi_0$. Since they are close in energy, we check for robustness in terms of their real-space localization, see Fig.\ \ref{fig:density}. 

To do this, we need to employ measures to quantize the distance and overlap between two states. Let $\ket{\textbf{r}_i, \sigma}$ denote the basis of sites, where $\textbf{r}_i = x_i \hat{x} + y_i \hat{y}$ denotes the location of the site measured with respect to the flux-tube. We can then define a density overlap measure between two states $\ket{\alpha}$ and $\ket{\beta}$ by
\begin{equation}\label{eq:overlap_measure}
    O(\ket{\alpha}, \ket{\beta}) = \sum_{i, \sigma} \min\left( \abs{\langle\textbf{r}_i, \sigma \mid \alpha \rangle}^2, \abs{\langle\textbf{r}_i, \sigma \mid \beta \rangle}^2\right).
\end{equation}
It is defined to yield values in the interval $[0,1]$ where $0$ implies no overlap at all while $1$ implies complete overlap, such that any state has perfect overlap with itself $O(\ket{\alpha}, \ket{\alpha})=1$. 

We can furthermore define two operators measuring the $x$- and $y$-localization respectively defined by
\begin{equation}
    X\ket{\textbf{r}_i, \sigma} = \frac{\abs{x_i}}{N_x}\ket{\textbf{r}_i, \sigma}, \qquad Y\ket{\textbf{r}_i, \sigma} = \frac{\abs{y_i}}{N_y}\ket{\textbf{r}_i, \sigma} 
\end{equation}
where $N_x = (L_x-1)/2$ and $N_y = (L_y-1)/2$ are normalizations that ensure expectation values always lie within the interval $[0,1]$ given by
\begin{equation}
    \langle X \rangle = \bra{\alpha} X \ket{\alpha} = \sum_{i, \sigma} \frac{\abs{x_i}}{N_x} \ \abs{\langle\textbf{r}_i, \sigma \mid \alpha \rangle}^2.
\end{equation}
Here $\langle X \rangle, \langle Y \rangle = 1$ denotes complete support on the farthest sites. 

We apply these measures on the two states by the crossing during adiabatic charge transport. The measures are then computed throughout the smooth decrease in system width to see how they may change. For the topological gap, this yields the results seen in Fig.\ \ref{fig:real_space}. 

For the case of open and periodic boundary conditions we find that the overlap decreases for larger values of $L_x$ (Fig.\ \ref{fig:op_overlap}). Furthermore, for the $x-$ and $y$-localization, one state stays close to the flux-tube, while the other moves further from the flux-tube as $L_x$ increases (Figs.\ \ref{fig:op_x}, \ref{fig:op_y}). This is in accordance with expectation as the crossing concerns a bulk state and an edge state. From the measures, we see that robustness increases with $L_x$ due to the larger spatial distance between the states. 

If we apply the measures to the case of only open boundary conditions, we find the same results. The only difference is a sudden spike in the $y$-localization for the edge state (Fig.\ \ref{fig:oo_y}). This is due to the state hybridizing with the zero energy states from the non-trivial Wilson topology as seen in Fig.\ \ref{fig:oo_charge}. As these states lie by the $y$-edges they create a sharp increase in $\langle Y \rangle$.

\section{Conclusion}\label{sec: Conclusion}

The considered system can exhibit non-trivial topological signatures with respect to both the one- and two-dimensional topological measures. This differentiates the system from usual topological insulators as it needs two topological invariants to describe the phases of matter. 

Usually if a two-dimensional system has non-trivial topology, opening the boundaries in one direction implies the existence of zero-energy edge states. The edges are thus conducting. This is the essence of bulk-boundary correspondence. However, the small width allows for gap openings such that we may describe the thin system as another topological insulator. We have shown that these gap openings occur due to hybridization between the edge states when the width $L_x$ is small enough. 

We find that the robustness of the one- and two-dimensional topologies are inversely related. The closer we move towards a given spatial dimension, the more prominent and robust the corresponding topological signature is. Using the smooth modulation of system width we can track gaps which do not close throughout the procedure. This means that we can create a smooth transition from a two-dimensional to a one-dimensional topological insulator. This can be done for gaps with both trivial and non-trivial one-dimensional topology. In between the limits of one- and two-dimensional topological insulators we get systems exhibiting both of the topological signatures. 

While bulk-boundary correspondence of the topological state studied here, protected by the finite hybridization gap, is currently not known to exhibit signatures distinguishing it from that of a one-dimensional topological state, it is anticipated that distinctions will be identified given the bulk-boundary correspondence of higher-dimensional finite-size topological phases~\cite{calderon2023_fst, pacholski2023, pal2024_fstwsm}. The topological invariant introduced to characterise topology of individual unit cells of the multiplicative Chern insulator~\cite{banerjee2025mci} and Bernevig-Hughes-Zhang model~\cite{ay2024signaturesquantumskyrmionhall}, motivated on first principles within the framework of the quantum skyrmion Hall effect~\cite{PhysRevB.109.155123, patil2024, patil2025microscopicfieldtheoriesquantum} by extension of matrix Chern-Simons theory~\cite{susskind2001quantum, polychronakos2001quantum}, is expected to be useful for this purpose.

\begin{acknowledgments}
    The work presented in this article is supported by Novo Nordisk Foundation grant NNF23OC0086670.
\end{acknowledgments}

\appendix

\section{Effective surface Hamiltonian} \label{app_sec_A}
We start from the following Bloch Hamiltonian of the Chern insulator, which describes a semi-infinite system in the $x$ direction with periodic boundary conditions along $y$:
\begin{align}
    H_{\textrm{op},L}(k_y) & =   \sum_{x=1}^{\infty} \ket{x}\bra{x} \otimes h(k_y)  \\
    & + \sum_{x = 1}^{\infty}\ket{x}\bra{x+1} \otimes (-t\sigma_z-i\Delta \sigma_y) + h.c. \notag
\end{align}
where $h(k_y)$ is given by
\begin{equation}
    h(k_y) = [M-2t\cos(k_y)]\sigma_z +2\Delta \sin(k_y) \sigma_x.
\end{equation}
Consider the Hamiltonian at the TRIM point in $k_y$, i.e.\ $k_y=q$, with $q=0,\pi$,
\begin{align}
    H_q & =   \sum_{x=1}^{\infty} \ket{x}\bra{x} \otimes M_q \sigma_z  \\
    & + \sum_{x = 1}^{\infty}\ket{x}\bra{x+1} \otimes (-t\sigma_z-i\Delta \sigma_y) + h.c. \notag
\end{align}
where $M_q=M-2t\cos(q)$. The above Hamiltonian is chiral and anticommutes with $\sigma_x$, so we can take the following ans{\"a}tze for zero energy states,
\begin{multline}
\ket{\Psi_0}\propto \ket{\sigma}_x\sum_x\zeta^x\ket{x}, \quad \sigma \in \{\pm\}, \quad \sigma_x\ket{\pm}_x=\pm\ket{\pm}_x,\\
 \ket{+}_x=\frac{1}{\sqrt{2}}
\begin{pmatrix}
1\\
1
\end{pmatrix}
,
\quad \ket{-}_x=\frac{1}{\sqrt{2}}
\begin{pmatrix}
1\\
-1
\end{pmatrix}
,
\end{multline}
where $\zeta$ denotes some constant to be determined. To solve the eigenvalue problem, $H_q \ket{\Psi_0}=0$, we get the following equation for the bulk,
\begin{equation}
M_q\zeta^x-t(\zeta^{x+1}+\zeta^{x-1})-\sigma\Delta(\zeta^{x+1}-\zeta^{x-1})=0,\\
\label{Eq:chareqn}
\end{equation}
which has the solutions
\begin{equation}
    \zeta_{\pm}(\sigma)=\frac{M_q\pm\sqrt{M_q^2-4(t^2-\Delta^2)}}{2(t+\sigma\Delta)}.
\end{equation}
Since $\zeta_{\pm}(\sigma=-1)=\zeta_{\mp}(\sigma=+1)^{-1}$ we define $\alpha_{\pm} = \zeta_{\pm}(\sigma=+1)$. From this we get the left edge state
\begin{equation}
\ket{\Psi_L}= \ket{+}_x\sum_x\psi_L(x)\ket{x}, \quad \psi_L(x) \propto \alpha_+^x-\alpha_-^x,
\end{equation}
where we have imposed the boundary condition $\psi_L(x=0)=0$. 

By considering a semi-infinite system $H_{\textrm{op},R}(k_y)$, which has an edge at $x=L_x$ and extends to infinity in the negative $x$ direction, we similarly derive the right edge state
\begin{multline}
\ket{\Psi_R}= \ket{-}_x\sum_x\psi_R(x)\ket{x}, \\ \psi_R(x)\propto \alpha_+^{L_x+1-x}-\alpha_-^{L_x+1-x},
\end{multline}
where we have imposed the boundary condition $\psi_R(x=L_x+1)=0$. Note that since $\abs{\alpha_\pm}<1$, we know that $\ket{\Psi_L}$ and $\ket{\Psi_R}$ are localized on the left and right edge respectively. 

\begin{figure}[h!]
    \includegraphics[width=0.45\textwidth]{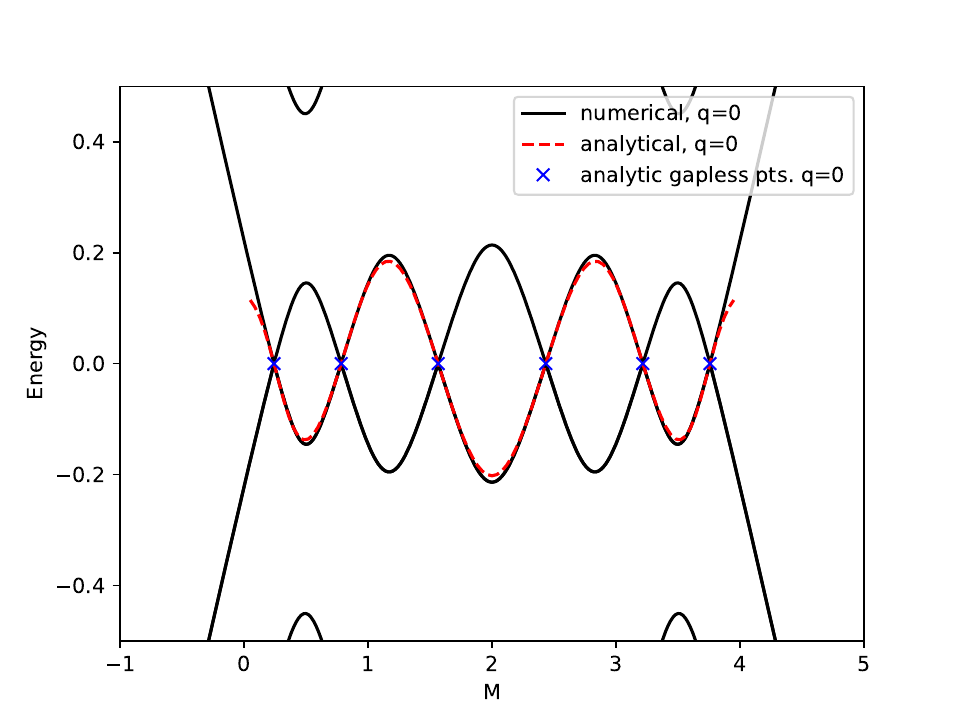}
    \caption{Energy vs.\ $M$ for $L_x=6$, $t=1$, $\Delta=0.22$ and $k_y=0$. The solid black lines represent the numerical spectra and the red dotted line is $\delta$ plotted according to Eq.\ (\ref{delta}). The blue crosses show the points in $M$ where the edge remains gapless according to Eq.\ (\ref{Eq:Mexpr}).}
    \label{fig:fciEvsM}
\end{figure}

We now consider the Hamiltonian for the finite system of width $L_x$, i.e.\
\begin{multline}
H_{\textrm{op}}(k_y)=\sum_{x=1}^{L_x}|x\rangle \langle x| \, H_{\textrm{op},L}(k_y) \, \sum_{x=1}^{L_x}|x\rangle \langle x|\\
=\sum_{x=1}^{L_x}|x\rangle \langle x| \, H_{\textrm{op},R}(k_y) \, \sum_{x=1}^{L_x}|x\rangle \langle x|,
\end{multline}
and compute its matrix elements within the space of the left and right edge states near $q=0$. The terms on the diagonal are
\begin{equation}
\begin{split}
H_L =& \braket{\Psi_L|H_{\textrm{op}}(k_y)|\Psi_L} = 2\Delta k_y,\\ 
H_R =& \braket{\Psi_R|H_{\textrm{op}}(k_y)|\Psi_R} = -2\Delta k_y.
\end{split}
\end{equation}
To compute the hybridization between the left and right edges, we first compute
\begin{widetext}
\begin{multline}
\sum_{x=1}^{L_x}|x\rangle\langle x| \, H_q \, \sum_{x=1}^{L_x}|x\rangle\langle x| \ket{\Psi_{L}} 
= \left[\sum_{x=1}^{L_x} \ket{x}\bra{x} \otimes M_q \sigma_z + \sum_{x = 1}^{L_x-1}\ket{x}\bra{x+1} \otimes (-t\sigma_z-i\Delta \sigma_y) + h.c \right] \sum_x\psi_L(x)\ket{x}\ket{+}_x \\
= \sum_{x=1}^{L_x}  M_q \psi_L(x) \ket{x}\ket{-}_x +\sum_{x=1}^{L_x-1} (-t-\Delta)\psi_L(x+1) \ket{x}\ket{-}_x+\sum_{x=2}^{L_x} (-t+\Delta)\psi_L(x-1) \ket{x}\ket{-}_x \\
= \sum_{x=1}^{L_x} [M_q \psi_L(x)-t(\psi_L(x+1)+\psi_L(x-1))-\Delta(\psi_L(x+1)-\psi_L(x-1))]\ket{x}\ket{-}_x \\
 + (t-\Delta)\psi_L(0)\ket{1}\ket{-}_x + (t+\Delta)\psi_L(L_x+1)\ket{L_x}\ket{-}_x \\
= (t+\Delta)\psi_L(L_x+1)\ket{L_x}\ket{-}_x,
\end{multline}
\end{widetext}
where we applied Eq.\ (\ref{Eq:chareqn}) as well as $\psi_L(0)=0$ in the final line. One sees that in the limit of $L_x \rightarrow \infty$, we have $H_q \ket{\Psi_{L}} = 0$ as it should be. The off-diagonal matrix element is then given by 
\begin{equation}\label{delta}
    \braket{\Psi_R|H_0|\Psi_L} = (t+\Delta)\psi_R^*(L_x)\psi_L(L_x+1) = \delta.
\end{equation}
With an appropriate choice of the global phases of the states, $\delta$ is real. We therefore denote $\delta \in \mathbb{R}$ as the magnitude of hybridization between the left and right edges. Denoting the left-right edge state basis by the Pauli matrix degree of freedom, $\nu$, we get the effective surface Hamiltonian for small $k_y$ as follows:
\begin{equation}
H_{\text{eff}}(k_y)=2\Delta k_y\nu_z+\delta\nu_x.
\end{equation}
The eigenvalues of the effective Hamiltonian for $k_y=0$ are thus $\pm\delta$. We show the agreement between numerics and this approximate, analytical result in Fig.\ \ref{fig:fciEvsM}.

From the expression of $\delta$, we observe that the hybridization vanishes when $\psi_L(L_x+1)=0$. The roots of this boundary condition provide us with the values of $M$ where the edge state remains gapless at $k_y=q$. Writing $\alpha_\pm$ as a complex number,
\begin{align}
\alpha_\pm&=Re^{i\theta},\\ 
R=\bigg{(}\frac{t-\Delta}{t+\Delta}\bigg{)}^{\frac{1}{2}},&\quad \cos\theta = \frac{M_q}{2\sqrt{t^2-\Delta^2}},
\end{align}
the left and right edge wavefunctions are given by,
\begin{align}
\psi_L(x)&=\frac{1}{\sqrt{N}}R^x\sin(\theta x),\\ \psi_R(x)&=\frac{1}{\sqrt{N}}R^{L_x+1-x}\sin[(L_x+1-x)\theta],\\ N&=\sum_{j=1}^{L_x}R^{2j}\sin^2(\theta j),
\end{align}
where $N$ denotes the normalization. In terms of $R$ and $\theta$, we have
\begin{equation}
\delta = \frac{(t+\Delta)}{N}R^{L_x+2}\sin(\theta)\sin[(L_x+1)\theta].
\label{Eq:deltaexpr}
\end{equation}
Hence $\delta = 0$ for
\begin{equation}
\theta = \frac{n\pi}{L_x+1},\quad (n=1,...,L_x).
\end{equation}
The hybridization thus vanishes for
\begin{equation}
M=2\sqrt{t^2-\Delta^2}\cos\bigg{(}\frac{n\pi}{L_x+1}\bigg{)}+2t
\label{Eq:Mexpr}
\end{equation}
with $n=1,...,L_x$.

\bibliography{apssamp}

\end{document}